\documentstyle[wstwocl]{article}
\begin{document}

\bibliographystyle{unsrt}    
\renewcommand{\a}{\alpha}
\renewcommand{\b}{\beta}
\renewcommand{\c}{\gamma}
\renewcommand{\d}{\delta}
\newcommand{\th}{\theta}
\newcommand{\TH}{\Theta}
\newcommand{\pa}{\partial}
\newcommand{\g}{\gamma}
\newcommand{\G}{\Gamma}
\newcommand{\A}{\Alpha}
\newcommand{\B}{\Beta}
\newcommand{\D}{\Delta}
\newcommand{\e}{\epsilon}
\newcommand{\E}{\Epsilon}
\newcommand{\z}{\zeta}
\newcommand{\Z}{\Zeta}
\newcommand{\k}{\kappa}
\newcommand{\K}{\Kappa}
\renewcommand{\l}{\lambda}
\renewcommand{\L}{\Lambda}
\newcommand{\m}{\mu}
\newcommand{\M}{\Mu}
\newcommand{\n}{\nu}
\newcommand{\N}{\Nu}
\newcommand{\x}{\chi}
\newcommand{\X}{\Chi}
\newcommand{\p}{\pi}
\newcommand{\r}{\rho}
\newcommand{\R}{\Rho}
\newcommand{\s}{\sigma}
\renewcommand{\S}{\Sigma}
\renewcommand{\t}{\tau}
\newcommand{\T}{\Tau}
\newcommand{\y}{\upsilon}
\newcommand{\Y}{\upsilon}
\renewcommand{\o}{\omega}
\newcommand{\q}{\theta}
\newcommand{\h}{\eta}
\def\ft#1#2{{\textstyle{{#1}\over{#2}}}}
\def\del{\partial}
\def\modb{\bar \Phi}
\def\mat{C}
\def\matb{\bar C}
\def\mpl{M_{\rm Pl}}

\newcommand{\st}{\scriptstyle}
\newcommand{\sst}{\scriptscriptstyle}
\newcommand{\mco}{\multicolumn}
\newcommand{\epp}{\epsilon^{\prime}}
\newcommand{\vep}{\varepsilon}
\newcommand{\ra}{\rightarrow}
\newcommand{\ab}{\bar{\alpha}}
\def\be{\begin{equation}}
\def\ee{\end{equation}}
\def\bea{\begin{eqnarray}}
\def\eea{\end{eqnarray}}

\def\jmp#1#2#3  {{\em J. Math. Phys.} {\bf#1} (#2) #3.}
\def\ijmp#1#2#3 {{\em Int. J. Mod. Phys.} {\bf#1} (#2) #3.}
\def\mpl#1#2#3  {{\em Mod. Phys. Lett.} {\bf#1} (#2) #3.}
\def\np#1#2#3   {{\em Nucl. Phys.} {\bf#1} (#2) #3.}
\def\pl#1#2#3   {{\em Phys. Lett.} {\bf#1} (#2) #3.}
\def\prep#1#2#3 {{\em Phys. Rep.} {\bf#1} (#2) #3.}
\def\prev#1#2#3 {{\em Phys. Rev.} {\bf#1} (#2) #3.}
\def\prl#1#2#3  {{\em Phys. Rev. Lett.} {\bf#1} (#2) #3.}

\setcounter{secnumdepth}{2} 

\begin{titlepage}
\begin{flushright} THU-95/30 \\ hep-th/9511019
\end{flushright}
\vfill
\begin{center}
{\large\bf SPECIAL GEOMETRY AND PERTURBATIVE ANALYSIS OF $N=2$}\\
{\large\bf HETEROTIC VACUA${}^\dagger$}\\
\vskip 7.mm
{B. de Wit }\\
\vskip 0.1cm
{\em Institute for Theoretical Physics} \\
{\em Utrecht University}\\
{\em Princetonplein 5, 3508 TA Utrecht, The Netherlands} \\
\end{center}
\vfill
\begin{center}
{\bf ABSTRACT}
\end{center}
\begin{quote}
The requirement of target-space duality and the use of
nonrenormalization theorems lead to strong constraints on the
perturbative prepotential that encodes the low-energy effective
action of $N=2$ heterotic superstring vacua. The analysis is
done in the context of special geometry, which governs the
couplings of the vector multiplets. The presentation is kept at
an introductory level. %
\vfill      \hrule width 5.cm
\vskip 2.mm
{\small\small
\noindent $^\dagger$ Based on an invited lecture at the
International Europhysics Conference on High-Energy Physics,
Brussels, July 27 - August 2, 1995; to be published in the proceedings.}
\end{quote}
\begin{flushleft}
October 1995
\end{flushleft}
\end{titlepage}


\title{SPECIAL GEOMETRY AND PERTURBATIVE ANALYSIS OF $N=2$
HETEROTIC VACUA}

\firstauthors{B. de Wit }

\firstaddress{Institute for Theoretical Physics, Utrecht
University,  Princetonplein 5, 3508 TA Utrecht, The
Netherlands}

%
%
%

\def\secondaddress

\twocolumn[\maketitle\abstracts{
The requirement of target-space duality and the use of
nonrenormalization theorems lead to strong constraints on the
perturbative prepotential that encodes the low-energy effective
action of $N=2$ heterotic superstring vacua. The analysis is
done in the context of special geometry, which governs the
couplings of the vector multiplets. The presentation is kept at
an introductory level. %
}]

\section{Special geometry}
Special geometry refers to the target-space geometry of $N=2$
supersymmetric vector multiplets, possibly
coupled to supergravity\cite{DWVP}. The physical states of a vector
multiplet are described by gauge fields $W^I_\mu$, doublets of
Majorana spinors $\Omega^I_i$ and complex scalars $X^I$. The
kinetic term for the scalars is a nonlinear sigma model which
defines the metric of the target space, the space parametrized by
the scalar fields.
Special geometry has the following characteristic features. The
Lagrangian is encoded in a holomorphic prepotential $F(X)$. In rigid
supersymmetry the fields $X^I$ can be regarded as independent
coordinates ($I=A=1,\ldots,n$). In the local case there is one
extra vector multiplet labeled by $I=0$, which provides the
graviphoton, but the $n+1$ fields $X^I$ are parametrized in terms
of $n$ holomorphic coordinates $z^A$. Often one chooses
{\it special} coordinates defined by $z^A= X^A/X^0$.
The target space is K\"ahlerian and the K\"ahler potential is given
by (the subscripts on $F$ denote differentiation)
\begin{eqnarray}
K(X,\bar X) &=& -i \bar X^AF_A(X) +i X^A \bar F_A(\bar X)\,,
\nonumber \\
K(z,\bar z) &=& -\log (-i \bar X^IF_I +i X^I \bar F_I)\,,
\end{eqnarray}
for rigid and for local supersymmetry, respectively. In the latter
case, the $2n+2$ quantities $(X^I,F_I)$ are parametrized by $n$
complex coordinates $z^A$. In more mathematical terms they define holomorphic
symplectic sections. The ensueing metric satisfies the following
curvature relations
\begin{eqnarray}
R^A{}_{\!BC}{}^{\!D} &=& - W_{BCE}\bar W^{EAD}\,, \nonumber \\
R^A{}_{\!BC}{}^{\!D} &=& 2\delta^A_{(B}\delta^D_{C)} - e^{2K}
W_{BCE}\bar W^{EAD}\,,
\end{eqnarray}
respectively, for the two cases. Here the tensor $W$ is related
to the third derivative of $F(X)$. Special geometry is also the geometry of
the moduli of Calabi-Yau spaces. This intriguing connection can be
understood in the context of type-II superstrings, whose
compactification on Calabi-Yau manifolds leads to
four-dimensional low-energy field theories with local $N=2$
supersymmetry.

The bosonic  kinetic terms read
\begin{eqnarray}
{\cal L} &=& {\textstyle{i\over 4\pi}}\Big( D_\mu F_I \,D^\mu \bar X^I -
D_\mu X^I \,D^\mu \bar F_I\Big) \label{vlagr}\\
&& -{\textstyle{i\over 16\pi}}
\Big( {\cal N}_{IJ}\,F_{\mu\nu}^{+I}F^{+\mu\nu J}\
-\ \bar{\cal N}_{IJ}\,F_{\mu\nu}^{-I} F^{-\mu\nu J} \Big)\,,
\nonumber
\end{eqnarray}
where $F^{\pm I}_{\m\n}$ denote the selfdual and anti-selfdual
field-strength components, and\footnote{%
  In the rigid case, $\cal N$ consists of only the first term and
  the $I=0$ component is suppressed.
  In general, $\cal N$ is complex. Its imaginary part
  is related to the gauge coupling constant, its real part to a
  generalization of the $\theta$ angle.}
\begin{equation}
{\cal N}_{IJ}=\bar
F_{IJ}+2i {{\rm Im}(F_{IK})\,{\rm Im}(F_{JL})\,X^KX^L\over {\rm
Im}(F_{KL})\,X^KX^L} \,.
\label{Ndef}
\end{equation}
\section{Symplectic reparametrizations}
{}From the Lagrangian (\ref{vlagr}) one defines the tensors
\be
G^+_{\mu\nu I}={\cal N}_{IJ}F^{+J}_{\mu\nu}\,,\quad G^-_{\mu\nu
I}=\bar{\cal N}_{IJ}F^{-J}_{\mu\nu}\,, \label{defG}
\ee
so that the Bianchi identities and equations of motion
for the Abelian gauge fields can be
written as
\begin{equation}
\partial^\mu \big(F^{+I}_{\m\n} -F^{-I}_{\m\n}\big)
=0\,,\qquad
\partial^\mu \big(G_{\mu\nu I}^+ -G^-_{\m\n I}\big) =0\,.
\label{Maxwell}
\end{equation}
These are invariant under the transformation
\be
\pmatrix{F^{+I}_{\mu\nu}\cr  G^+_{\mu\nu I}\cr} \longrightarrow
\pmatrix{U&Z\cr W&V\cr} \pmatrix{F^{+I}_{\mu\nu}\cr  G^+_{\mu\nu
I}\cr}\label{FGdual}
\ee
where $U^I_{\,J}$, $V_I^{\,J}$, $W_{IJ}$ and $Z^{IJ}$ are
constant real  $(n+1)\times(n+1)$ submatrices.
The transformations for the anti-selfdual tensors follow by
complex conjugation. From (\ref{defG},\ref{FGdual}) one derives
that $\cal N$ transforms as
\begin{equation}
{\cal N}_{IJ} \longrightarrow (V_I{}^K {\cal N}_{KL}+ W_{IL} )\,
\big[(U+ Z{\cal N})^{-1}\big]^L{}_J  \,.\label{nchange}
\end{equation}
To ensure that $\cal N$ remains a symmetric tensor, at
least in the generic case, the transformation
(\ref{FGdual}) must be an element of $Sp(2n+2,{\bf R})$
(disregarding a uniform scale transformation).
The required change of $\cal N$ is induced by a change of the scalar
fields, implied by
\be
\pmatrix{X^{I}\cr  F_{I}\cr} \longrightarrow  \pmatrix{\tilde
X^I\cr\tilde F_I\cr}=
\pmatrix{U&Z\cr W&V\cr} \pmatrix{X^{I}\cr  F_I\cr}\,.
\label{transX}
\ee
In this transformation we include a change of $F_I$. Because the
transformation belongs to $Sp(2n+2,{\bf R})$, one can show that
the new quantities $\tilde F_I$ can be written as
the derivatives of a new function $\tilde F(\tilde X)$.
The new but equivalent set of equations of motion one obtains by
means of the symplectic transformation (properly extended to other
fields), follows from the Lagrangian based on $\tilde F$. In
special cases $F$ remains unchanged, $\tilde F(\tilde X) = F(\tilde
X)$, so that the theory is {\it invariant} under the
corresponding transformations.

The symplectic transformations (\ref{FGdual}) cause electric
fields to transform into magnetic fields and vice versa.  The
interchange of electric and magnetic fields is known as
electric-magnetic duality. Under the transformation
with $U=V=0$ and $W=-Z ={\bf 1}$, $F^{+I}_{\mu\nu}$ and
$G^+_{\mu\nu I}$ are simply interchanged, while $\cal N$
transforms into  $-{\cal N}^{-1}$. Since the coupling constants
are thus replaced by their inverses, electric-magnetic duality
relates the strong- and
weak-coupling description of the theory. Electric-magnetic duality is
a special case of so-called $S$ duality. The coupling constant
inversion is then part of an $Sl(2,{\bf Z})$ group. This
situation is known in the context of string theory and
lattice gauge theories. Other symplectic transformations (with
$Z$=0) can be discussed at the perturbative level and may involve a
shift of the generalized $\theta$ angles. In nonabelian
gauge theories $\theta$ is periodic, so that $\cal N$ is
defined up to the addition of certain discrete real constants.

\section{Semiclassical theory of monopoles and dyons}
To elucidate some important features of
the symplectic reparametrizations, let us discuss the effective
action of abelian gauge fields, possibly obtained from a
nonabelian theory by integrating out certain fields. We
write the matrix $\cal N$ in terms of generalized coupling
constants and $\theta$ angles, according to
\be
{\cal N}_{IJ}= {\theta_{IJ}\over 2\pi} -i {4\pi\over g^2_{IJ}} \,
.\label{caln}
\ee
This matrix can be compared to a generalization of the
permeability and permittivity that is conventionally used in the
treatment of electromagnetic fields in the presence of a medium. The fields
$G_{\mu\nu I}$ are thus generalizations of the displacement and
magnetic fields, while $F^I_{\mu\nu}$ corresponds to the electric
fields and magnetic inductions. So far we have considered an abelian
theory without charges. It is straightforward to introduce
electric charges by introducing an electric current in the
Lagrangian. However, to consider duality tranformations one must
also introduce magnetic currents into the field equations, so
that when electric fields tranform into magnetic fields and vice
versa, the electric and magnetic currents transform accordingly. The
magnetic currents occur as sources in the Bianchi identity and
describe magnetic monopoles.

Electric and magnetic charges are conveniently defined in terms
of flux integrals over closed spatial surfaces that surround the
charged objects,
\bea
\oint_{\partial V} (F^+ + F^-)^I \!&=& \! 2\pi \,q_{\rm m}^{I}\,,\nonumber \\
\oint_{\partial V} (G^+ + G^-)_I \!&=&  \! -2\pi \,q_{{\rm e}I} \,.
\label{charges}
\eea
With these definitions a static point charge at the origin
exhibits magnetic inductions and electric fields equal to $\vec
r/(4\pi r^3)$ times $2\pi q_{\rm m}^{I}$ and
$\ft12g^2(q_{{\rm e}I} + q_{\rm m}^J \,\theta_{IJ}/2\pi)$,
respectively. Note that $q_{\rm e}$ does not
coincide with the electric charge. From (\ref{charges}) it follows that the
charges must transform under symplectic rotations according to
\be
\pmatrix{q_{\rm m}^{I}\cr  -q_{{\rm e}I}\cr} \longrightarrow
\pmatrix{U&Z\cr W&V\cr} \pmatrix{q_{\rm m}^{I}\cr  -q_{{\rm
e}I}\cr} \,.
\ee

As is well known, the charges are subject to a generalized Dirac
quantization condition, due to
Schwinger and Zwanziger. To derive this condition, consider a test
particle with charges $q^{\prime}_{\rm e}$ and $q^{\prime}_{\rm
m}$ in the field of a
heavy dyon with charges $q_{\rm e}$ and $q_{\rm m}$ (for
simplicity we restrict ourselves to a single gauge field).
The equation
of motion of the test particle is assumed to be invariant under
duality transformations. There is only one
symplectic invariant that one can construct from the test particle
charges and the dyon fields, namely $q^{\prime}_{\rm m}G_{\mu\nu}
+q^{\prime}_{\rm e}F_{\mu\nu}$, where $F_{\m\n}$ and $G_{\m\n}$
represent the fields induced by (\ref{charges}).
Inserting this combination into the field equation of the test particle
yields a generalization of the Lorentz force,
\be
m{{\rm d}^2 x^\mu\over {\rm d}\tau^2} = \Big\{{\textstyle{1\over
2}} \Big(q_{\rm e}^\prime+{\theta\over 2\pi} q_{\rm m}^\prime
\Big) F^{\m\n} + {\textstyle{1\over 2}} q_{\rm
m}^\prime{4\pi\over g^2}   {}^\ast\!F^{\m\n}  \Big\} {{\rm
d}x_\nu\over {\rm d}\tau}  \,.
\label{particleeq}
\ee
The angular momentum $\vec L= m\vec r\times
\dot{\vec r}$ of the test particle is
not invariant in the dyon field (taken at the origin) and one
must include the contribution of the electromagnetic fields. The
total angular momentum vector,
\be
\vec J =\vec L +{\vec r\over r} {q_{\rm e}q^\prime_{\rm m}-q_{\rm
m}q^\prime_{\rm e}\over 4}\,,
\ee
is indeed a constant of the motion.
Quantum-mechanically the component of this vector along $\vec r$
must be an integer times $\hbar/2$, so that one obtains a
quantization condition for $q_{\rm e}q^\prime_{\rm m}-q_{\rm
m}q^\prime_{\rm e}$. It implies that the allowed
electric and magnetic charges comprise a lattice such that surface
elements spanned by the lattice vectors are equal to a
multiple of the Dirac unit $2\hbar$, as shown in fig. 1. In
addition, the lattice should be consistent with the
periodicity of the $\theta$ angle%
\footnote{%
   The normalization of the $\theta$ angle is
   fixed by the assumption that instantons yield an integer value
   for the Pontryagin index $(32\pi^2)^{-1}\int{\rm d}^4 x
   \;{}^\ast\!F F$ in a nonabelian extension of the theory.%
}, $\theta\to \theta + 2\pi$ corresponding to ${\cal
N} \to {\cal N} + {\bf 1}$. This shift is associated with a symplectic
transformation with $U=V=W={\bf 1}$ and $Z=0$, so that the
charges transform according to $q_{\rm e}\to q_{\rm e}+ q_{\rm m}$,
while $q_{\rm m}$ remains invariant. This transformation is
contained in the discrete subgroup $Sp(2n+2,{\bf Z})$ that leaves
the charge lattice invariant.
\setlength{\unitlength}{0.5mm}
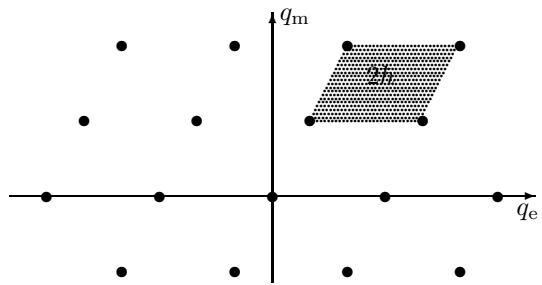
\begin{figure}[t]
\begin{picture}(130,70)(-15,0)
\put(0,20){\vector(1,0){140}}
\put(135,15){$q_{\rm e}$}
\put(70,-3){\vector(0,1){72}}
\put(72,67){$q_{\rm m}$}
\multiput(10,20)(30,0){5}{\circle*{3}}
\multiput(30,0)(30,0){4}{\circle*{3}}
\multiput(20,40)(30,0){4}{\circle*{3}}
\multiput(30,60)(30,0){4}{\circle*{3}}
\multiput(80,40)(1,0) {30}{\circle*{.1}}
\multiput(80.5,41)(1,0) {30}{\circle*{.1}}
\multiput(81,42)(1,0) {30}{\circle*{.1}}
\multiput(81.5,43)(1,0) {30}{\circle*{.1}}
\multiput(82,44)(1,0) {30}{\circle*{.1}}
\multiput(82.5,45)(1,0) {30}{\circle*{.1}}
\multiput(83,46)(1,0) {30}{\circle*{.1}}
\multiput(83.5,47)(1,0) {30}{\circle*{.1}}
\multiput(84,48)(1,0) {30}{\circle*{.1}}
\multiput(84.5,49)(1,0) {30}{\circle*{.1}}
\put(95,50){$2\hbar$}
\multiput(85,50)(1,0) {30}{\circle*{.1}}
\multiput(85.5,51)(1,0) {30}{\circle*{.1}}
\multiput(86,52)(1,0) {30}{\circle*{.1}}
\multiput(86.5,53)(1,0) {30}{\circle*{.1}}
\multiput(87,54)(1,0) {30}{\circle*{.1}}
\multiput(87.5,55)(1,0) {30}{\circle*{.1}}
\multiput(88,56)(1,0) {30}{\circle*{.1}}
\multiput(88.5,57)(1,0) {30}{\circle*{.1}}
\multiput(89,58)(1,0) {30}{\circle*{.1}}
\multiput(89.5,59)(1,0) {30}{\circle*{.1}}
\multiput(90,60)(1,0) {30}{\circle*{.1}}
\end{picture}
\caption{Lattice of `electric' and magnetic charges}
\vspace{-4mm}
\end{figure}

\section{Wilsonian action and nonrenormalization}
We now elucidate some consequences of special geometry in
the example of $SU(2)$ $N=2$
supersymmetric gauge theory\cite{SeibWit}. The classical theory
is described by $F_{\rm class}(\vec X)= {\textstyle{i\over 2}}S\,
\vec X^2$, where $\vec X$ is an $SU(2)$ vector and $iS=
\theta/2\pi +i 4\pi/g^2$.
In the context of heterotic string theory $S$ is
the dilaton field, which is the scalar component of a vector
multiplet. We momentarily restrict this multiplet to be constant, which
preserves supersymmetry. The
nonabelian theory has a potential with flat valleys whenever the
real and imiginary parts of $\vec X$ are
parallel, so that $SU(2)$ is broken down to $U(1)$. In the
Wilsonian action we integrate out the fields of momenta higher
than a certain scale $\Lambda$. Restricting ourselves to the
vector multiplet associated with the unbroken $U(1)$, the
Wilsonian effective action is then, by $N=2$ supersymmetry, encoded in a
{\it holomorphic} function $F_{\rm W}(X,S)$; because $S$ can be regarded
as the scalar component of a vector multiplet, holomorphicity
applies also to $S$. However, at the same time $S$ is a
loop-counting parameter. The one-loop result is therefore
$S$-independent and  explicit calculation shows that it is equal
to $(i/2\pi) X^2\ln (X^2/\L^2)$. The coeffient in
front of this expression is directly related to the one-loop beta
function and the chiral anomaly%
\footnote{%
   Including the one-loop correction the gauge coupling ${\cal N}$
   becomes equal to ${\cal N}_{\rm eff}= \theta/2\pi
   -i4\pi/g^2 -(i/\pi)(\ln (X^2/\L^2) +3)$ and satisfies
   $\L^2\partial/\partial \L^2 {\cal N}_{\rm eff} = (-i/\pi)[ -\ft{11}{12}
   + \ft26+ \ft 2{24} ]{\cal C}_2$, where the separate terms refer to the
   beta-function contribution from vectors, scalars and spinors,
   respectively; ${\cal C}_2$ is the second-order Casimir invariant, which
   equals 2 for $SU(2)$. Nonperturbatively $\cal N$ also receives
   real $\L$-dependent corrections, which lead to a
   renormalization of the $\theta$ angle. See ref.~2
   and references quoted therein.%
}. The latter is related to the fact that $X^{-2}F_{\rm W}(X,S)$
changes under a phase transformation $X\to e^{i\a}X$ by an
additive term $-\a/\pi$.

Now we invoke a nonrenormalization argument which hinges on the
fact that, perturbatively, the result should be independent of the
$\theta$ angle, as this parameter multiplies only the total
divergence ${}^\ast\!F F$ in the Lagrangian. However, the
requirement that the Wilsonian action should be independent of
$\theta$, while the corresponding prepotential should depend
holomorphically on $S$, excludes all perturbative corrections
beyond the one-loop level. Therefore we may write
\be
F_{\rm W}(X,S) = {i\over 2\pi} X^2\Big\{\ln {X^2\over \Lambda^2
e^{-\pi S}} + f^{\rm n.p.}\Big({\Lambda^4 e^{-2\pi S} \over
X^4}\Big)
\Big\}\,,       \label{fullF}
\ee
where the last term denotes the nonperturbative contributions.
These take a restricted
form. First of all, nonperturbatively, the action is
invariant under discrete shifts of $\theta$ equal to multiples of
$2\pi$. Secondly,
corrections from instantons break the invariance under phase
transformations $X\to e^{i\a}X$ to ${\bf Z}_4$. This is tied to
the 8 independent fermionic zero-modes that exist in the
instanton background (the number of zero-modes is related
to the one-loop axial anomaly coefficient through the
Atiyah-Singer index theorem). So we write the nonperturbative
corrections as a function of the dimensionless ${\bf Z}_4$
invariant $X^4/\L^4$. Thirdly, by assigning an extra
transformation to $S$, namely $S\to S-2i\a/\pi$, the chiral
anomaly cancels at one loop. This can be generalized beyond one
loop by properly adjusting the $S$-dependent subtractions, so
that the combined transformation of $X$ and $S$ will constitute
an exact invariance.
The explicit form of the function $f^{\rm n.p.}$
is of course subtraction dependent. The parametrization
(\ref{fullF}) is entirely in agreement with explicit instanton
calculations (note that the real part of $2\pi S$ equals the
one-instanton action $8\pi^2/g^2$) and exhibits the cut-off
dependence characteristic for supersymmetric gauge theories.

We draw attention to the fact that the function (\ref{fullF}) is
not single valued. Because of the (perturbative) logarithmic
correction, $F_{\rm W}$ is determined up to a quadratic function
$X^2$ with an integer coefficient. The logarithmic singularity is
due to the fact that the mass of the charged particles that we
integrated out, tends to zero when $|X|$ vanishes. Going around
the branch-cut by $X\to e^{i\pi} X$ is equivalent to a symplectic
transformation with $U=V=-1$, $W=2$ and $Z=0$. These monodromies
play a central role in understanding the groundstate structure
of $N=2$ supersymmetric Yang-Mills theories, as demonstrated by
Seiberg and Witten\cite{SeibWit}.
By exploiting symplectic reparametrizations they decribe similar
singularities in the nonperturbative domain.
Again these singularities can be understood as the result of
certain electrically charged states becoming massless, where
`electric' refers to the new basis obtained after applying the
symplectic reparametrization. In the original (semiclassical)
basis these states then correspond to dyons with nonzero magnetic
charge.

\section{$N=2$ Heterotic vacua}
Finally we consider the prepotential of vector multiplets relevant
for $N=2$ heterotic vacua. In such
vacua there are at least two abelian gauge fields, one associated
with the graviphoton and one with the dilaton field. (This can be
deduced from supergravity alone, provided the dilaton
is contained in a so-called vector-tensor multiplet.) In toroidal
compactifications there are two extra vector multiplets
associated with the complex toroidal moduli $T$ and $U$.
The classical prepotential is uniquely determined\cite{FVP} (up
to symplectic reparametrizations) by the requirement that the
dilaton couples universally at the string tree level and the
effective action does not depend on the $\theta$ angle,
\be
F_{\rm class}(X) = - {X^1\over X^0} \Big[X^2X^3 -
{\textstyle\sum_{I\geq 4}}
(X^I)^2\Big] \,.  \label{function}
\ee
This function corresponds to the product manifold $[SU(1,
1)/U(1)]\times[SO(2,n-1)/(SO(2)\times SO(n-1))]$. The $SU(1,
1)/U(1)$ coordinate is the dilaton field $iS=X^1/X^0$, whose real
part corresponds the string coupling constant; other
moduli are given by $iT=X^2/X^0$, $iU=X^3/X^0$,
etc. The objective is to consider the perturbative
corrections which, as above, originate
entirely from one-loop effects and cause an $S$-independent
addition to (\ref{function}). An immediate
problem is that the gauge couplings do not uniformly vanish in
the large dilaton limit. To set up string perturbation theory
consistently we must therefore change our basis by means of a symplectic
reparametrization. As it turns out, this reparametrization is
such that the prepotential $F$ no longer exists. Fortunately the
latter is merely a
technical problem\cite{Ceresole}. In the new basis the classical
Lagrangian is $SO(2,n-1)$ invariant.

The one-loop correction should be invariant under
target-space dualities, which are perturbative and expected to
leave the dilaton invariant.
Classically they coincide with $SO(2,n-1)$, but for finite string
coupling we expect only a discrete subgroup to be
relevant\cite{SEN}. Not surprisingly, in view of the high
symmetry of the classical result, there are no
modifications of (\ref{function}) that preserve the invariance
under the full $SO(2,n-1)$. Furthermore, we expect the
corrections to exhibit a similar lack of
single-valuedness as noted in the previous section. The corresponding
monodromy, whose identification depends on the proper choice for
the new symplectic basis, is induced by a
symplectic transformation that interferes
with the $SU(1,1)$ invariance of the dilaton.

It turns out that the one-loop contribution to the function
${\cal F}_{\rm W}\equiv  i(X^0)^{-2}F_{\rm W}$ must therefore be
invariant under target-space duality transformations, up to a
restricted polynomial of the moduli with discrete real
coefficients\cite{DWKLL}. For
example, in toroidal compactifications of six-dimensional
$N=1$ string vacua, where we have only $T$ and $U$, the
transformation $T\to (aT-ib)/(icT+d)$ with integer
parameters satisfying $ad-bc =1$, induces the following result on
the one-loop correction\cite{DWKLL,A},
\be
{\cal F}^{(1)}(T,U)\to (icT+d)^{-2} [{\cal F}^{(1)}(T,U)+\Xi(T,
U)] \,,
\ee
where $\Xi$ is a quadratic polynomial in the variables
$(1,iT,iU,TU)$. Hence $\partial_T^3\Xi=\partial_U^3\Xi=0$.
The appearance of $\Xi$ complicates the symmetry properties of
the one-loop moduli prepotential, which would otherwise be a
modular function of weight $-2$. It encodes the
monodromies at singular points in the moduli space (for instance,
at $T\approx U$) where one has an enhancement of the gauge symmetry.
Knowledge of these singularities allows one to
determine certain derivatives of ${\cal F}^{(1)}$ in terms of
standard modular functions. For a detailed discussion of the
monodromies in the toroidal case, we refer to refs. 8 and 9.

In the presence of the one-loop correction the dilaton field is no
longer invariant under target-space duality. This can be
understood from the fact that the dilaton belongs originally to a
vector-tensor multiplet and is only on-shell equivalent
to a vector multiplet. One can
always redefine $S$ such that it becomes invariant,
but then it can no longer be interpreted as the scalar component
of a vector multiplet\cite{DWKLL}. Interestingly enough, these perturbative
results have been confirmed by explicit calculations based on
`string duality'. For this we refer to A. Klemm's contribution to
these proceedings. \\[4mm]
This talk is based on work done in collaboration with V.
Kaplunovsky, J. Louis, D. L\"ust and A. Van Proeyen.
I thank M. Faux for stimulating conversations.
Further details and references to the literature
can be found in the papers listed below.




\end{document}